\documentclass{JINST}
\title{Muon detector for the COSINE-100 experiment}

\author{COSINE-100 Collaboration}
\author{H.~Prihtiadi$^{a,b}$, G.~Adhikari$^c$, P.~Adhikari$^c$, E.~Barbosa~de~Souza$^{d}$, N.~Carlin$^{e}$, S.~Choi$^{f}$, W.Q.~Choi$^{g}$, M.~Djamal$^{a}$, A.C.~Ezeribe$^{h}$,
C.~Ha$^{b}$, I.S.~Hahn$^{i}$, A.J.F.~Hubbard$^{d}$, E.J.~Jeon$^{b}$, J.H.~Jo$^{d}$, H.W.~Joo$^{f}$, W.~Kang$^{j}$, W.G.~Kang$^{b}$, M.~Kauer$^{k}$,
B.H.~Kim$^{b}$,H.~Kim$^{b}$, H.J.~Kim$^{l}$, K.W.~Kim$^{b}$, N.Y.~Kim$^{b}$, S.K.~Kim$^{f}$, Y.D.~Kim$^{b}$, Y.H.~Kim$^{b}$,
V.A.~Kudryavtsev$^{h}$, H.S.~Lee$^{b}$, J.~Lee$^{b}$, J.Y.~Lee$^{l}$, M.H.~Lee$^{b}$, D.S.~Leonard$^{b}$, K.E.~Lim$^{d}$, W.A.~Lynch$^{h}$, R.H.~Maruyama$^{d}$,
F.~Mouton$^{h}$, S.L.~Olsen$^{b}$, H.K.~Park$^{b}$, H.S.~Park$^{m}$, J.S.~Park$^{b}$, K.S.~Park$^{b}$, W.~Pettus$^{d}$, Z.P.~Pierpoint$^{d}$,
S.~Ra$^{b}$, F.R.~Rogers$^{d}$, C.~Rott$^{j}$, A.~Scarff$^{h}$, N.J.C.~Spooner$^{h}$, W.G.~Thompson$^{d}$, L.~Yang$^{n}$, and S.H.~Yong$^{b}$\\
\llap{$^a$} Department of Physics, Bandung Institute of Technology, Bandung 40132, Indonesia\\
\llap{$^b$} Center for Underground Physics, Institute for Basic Science~(IBS), Daejeon 34047, Republic of Korea\\  
\llap{$^c$} Department of Physics, Sejong University, Seoul 05006, Republic of Korea\\
\llap{$^d$} Department of Physics, Yale University, New Haven, CT 06520, USA\\
\llap{$^e$} Physics Institute, University of S\~{a}o Paulo, S\~{a}o Paulo 05508-090, Brazil\\
\llap{$^f$} Department of Physics and Astronomy, Seoul National University, Seoul 08826, Republic of Korea\\    
\llap{$^g$} Korea Institute of Science and Technology Information, Daejeon 34141, Republic of Korea\\    
\llap{$^h$} Department of Physics and Astronomy, University of Sheffield, Sheffield S3 7RH, United Kingdom\\    
\llap{$^i$} Department of Science Education, Ewha Womans University, Seoul 03760, Republic of Korea\\    
\llap{$^j$} Department of Physics, Sungkyunkwan University, Suwon 16419, Republic of Korea\\    
\llap{$^k$} Department of Physics and Wisconsin IceCube Particle Astrophysics Center, University of Wisconsin-Madison, Madison, WI 53706, USA\\    
\llap{$^l$} Department of Physics, Kyungpook National University, Daegu 41566, Republic of Korea\\    
\llap{$^m$} Korea Research Institute of Standards and Science, Daejeon 34113, Republic of Korea\\    
\llap{$^n$} Department of Physics, University of Illinois at Urbana-Champaign, Urbana, IL 61801, USA\\    
E-mail: \email{hyunsulee@ibs.re.kr, cha@ibs.re.kr}}

\abstract{ 
The COSINE-100 dark matter search experiment has started taking physics data with the goal of performing an independent measurement of the annual modulation signal observed by DAMA/LIBRA. A muon detector was constructed by using plastic scintillator panels in the outermost layer of the shield surrounding the COSINE-100 detector. It is used to detect cosmic ray muons in order to understand the impact of the muon annual modulation on dark matter analysis. Assembly and initial performance test of each module have been performed at a ground laboratory. The installation of the detector in Yangyang Underground Laboratory (Y2L) was completed in the summer of 2016. Using three months of data, the muon underground flux was measured to be 328 $\pm$~1(stat.)$\pm$~10(syst.)~muons/m$^2$/day. In this report, the assembly of the muon detector and the results from the analysis are presented. 
}


\begin{document}
\section{Introduction}

When a primary cosmic ray particle interacts with a molecule in the atmosphere, a shower of energetic hadrons including pions $(\pi^\pm)$ and kaons $(K^\pm)$ are generated. These quickly decay and produce relativistic muons~\cite{muon1, pdg} that are highly boosted with the livetime long enough to reach the surface and deep underground laboratories. Therefore, they can be a significant background source in underground physics experiments. 
Because muons can produce spallation neutrons that can produce scintillation signals in a crystal detector and could mimic those expected for dark matter particles. It is essential to understand muon and muon-induced events in direct dark matter search experiments~\cite{edelweissmuon,luxmuon}. 

Phenomena attributed to dark matter could be explained as being due to the effects of previously unseen particles such as weakly interacting massive particles (WIMPs) which are strongly motivated by theory~\cite{wp1,wp2}.
Numerous experiments have searched for WIMPs by directly detecting nuclei recoiling from WIMP--nucleon interactions in low background detectors located deep underground~\cite{dm1,dm2,dm3}. A positive annual modulation signal would be a signature for dark matter due to changes in the relative motion of the Earth orbits the Sun in the galaxy cluster. 
Among various dark matter search experiments, the DAMA/LIBRA experiment using a NaI(Tl) crystal array~\cite{DAMA_01,DAMA_02,DAMA_03} is extremely interesting because an annual modulation was observed, which can be interpreted as WIMP--nucleon interactions~\cite{savage}. 

Owing to temperature and density variations in the atmosphere, it is well known that the muon rate is modulated annually~\cite{muonmod1,dmice1,dmice2}. This fact motivated the consideration of muons as a possible source to explain the DAMA/LIBRA signal~\cite{nygren,jonathan,ralston}. 
However, this possibility remains unresolved because there is a phase difference between the muon and expectations for WIMP-induced modulations~\cite{Sea_mod}, and the rates for muon-induced neutrons predicted in simulation-based studies~\cite{ejjeon,vitaly} are too low to explain the modulation observed by the DAMA/LIBRA experiment.

The COSINE-100 experiment is a collaboration between the Korea Invisible Mass Search~(KIMS)~\cite{kimsnai1,kimsnai2,kimsnai3} and DM-Ice~\cite{dmice1,dmice2} experiments to confirm or refute the annual modulation signal observed by the DAMA/LIBRA experiment.
The COSINE-100 experiment was commissioned and has been taking physics data since September 2016 with a 106~kg array of low-background NaI(Tl) crystals in the Yangyang Underground Laboratory (Y2L) in Korea~\cite{cosine}. One of its features is the outermost plastic scintillator panels surrounding the main detector, which are used to tag muon events. In this paper, the assembly and initial performance of the muon detector is described as well as a measurement of the cosmic ray flux.

\section{Assembly of The Muon Detector}
\begin{figure}[htbp]	
	\centering
	\begin{tabular}{cc}
	\includegraphics[width = 0.35 \textwidth] {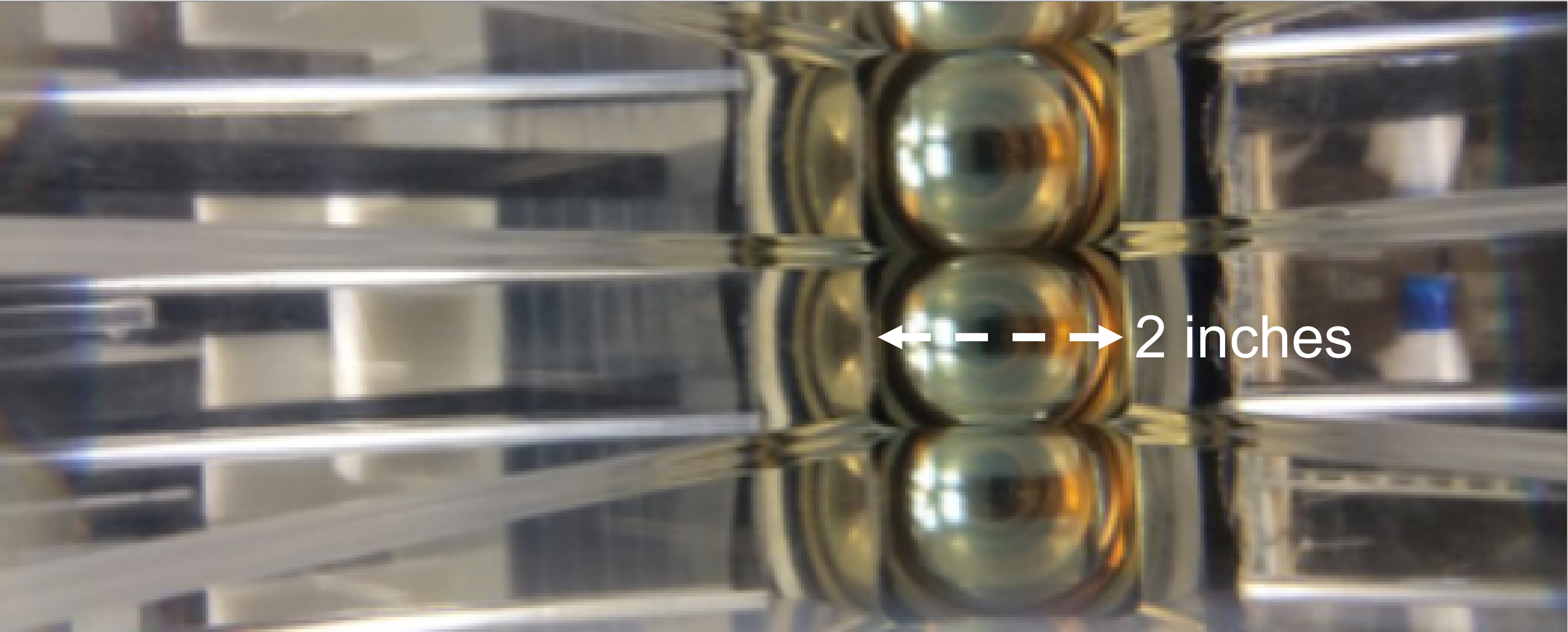}&
	\includegraphics[width = 0.55 \textwidth] {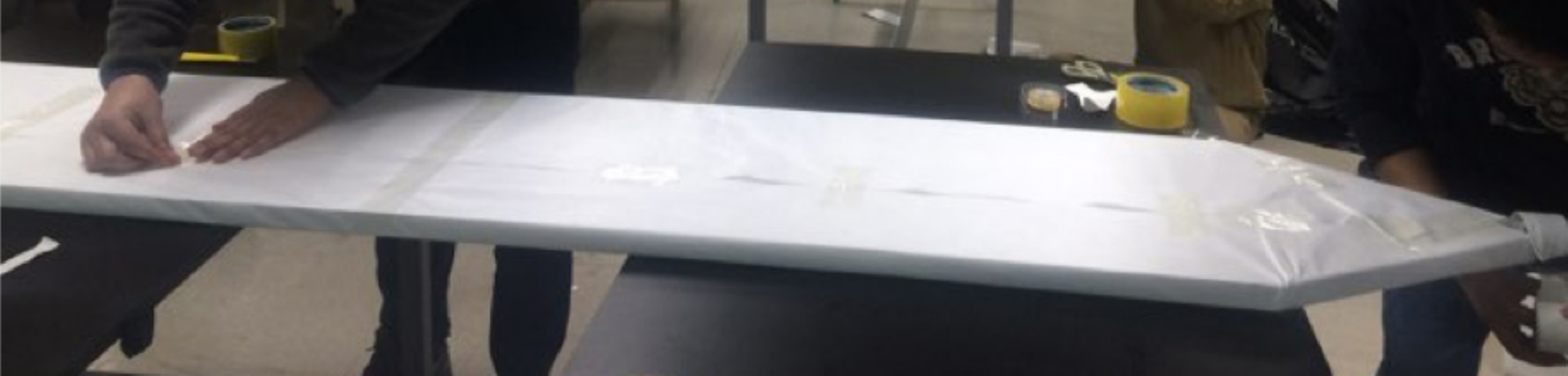}\\
	(a) & (b) \\
  \includegraphics[width=.4\textwidth]{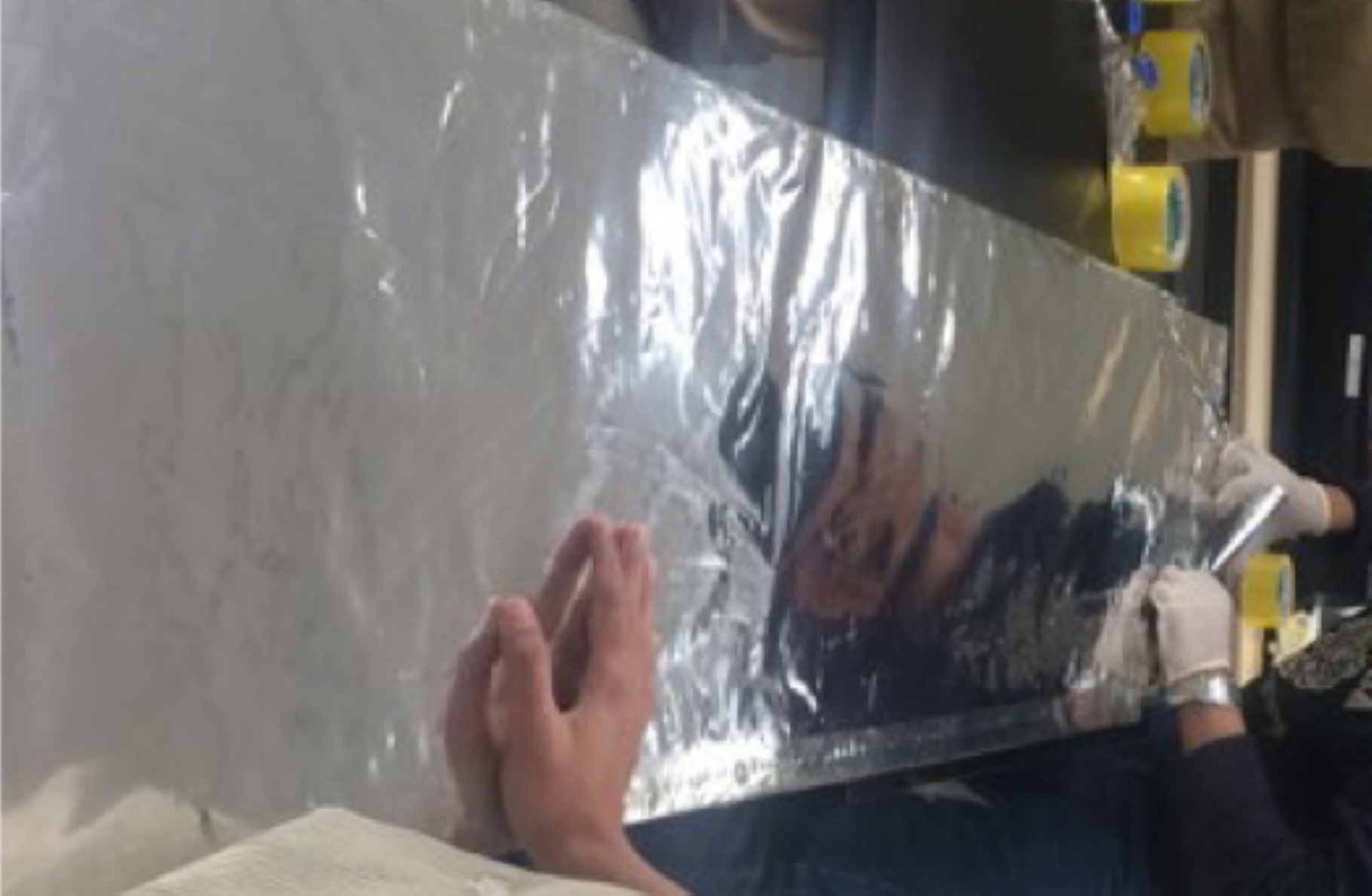} &
  \includegraphics[width=.58\textwidth]{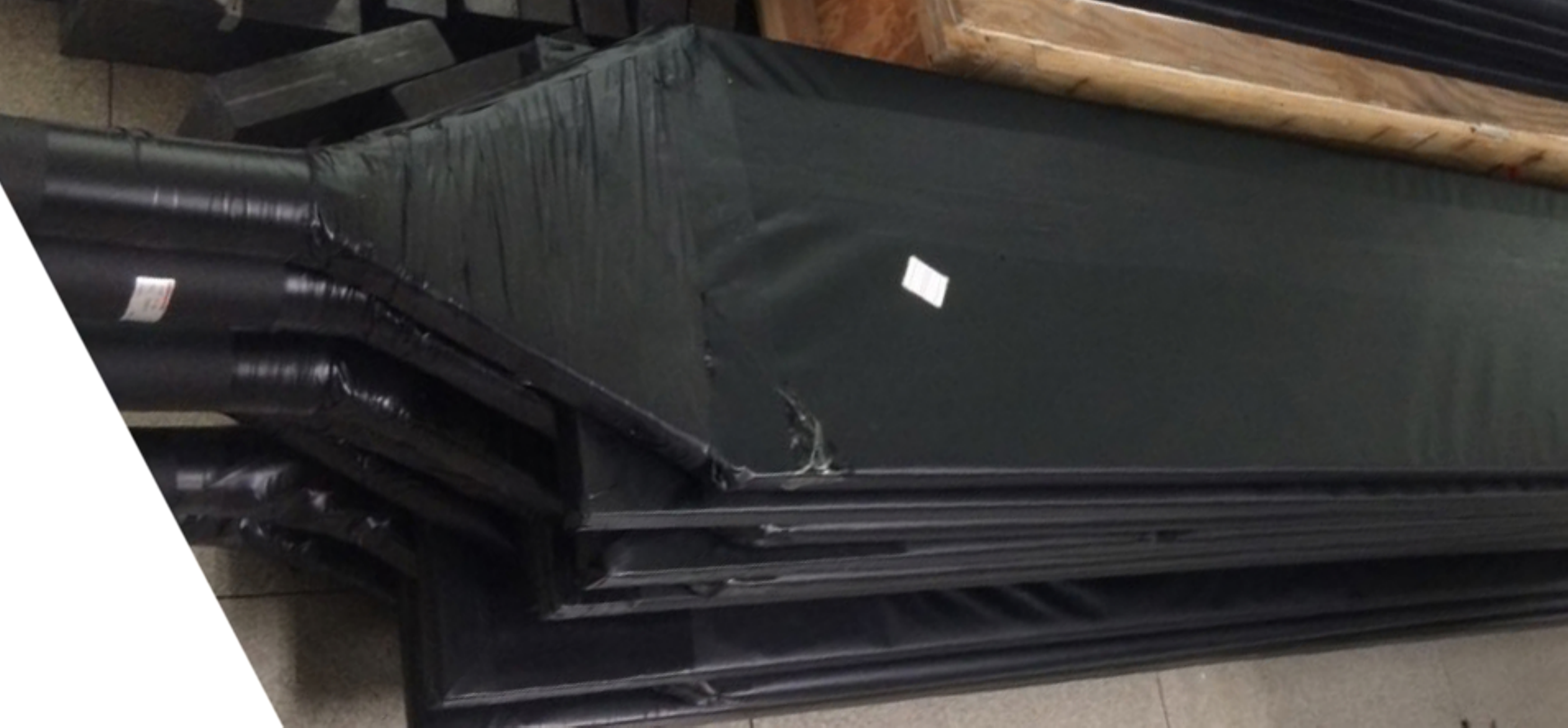} \\
	(c) & (d) \\
	\end{tabular}
	\caption{Assembly procedure of the plastic scintillator panels. (a) The light guide with PMT is attached with BC-600 optical cement. (b) A panel is being wrapped with a TYVEK reflector. (c) A panel is being wrapped with an aluminum foil. (d) Muon panels are wrapped with a black vinyl sheet.}
	\label{muonassemble}
\end{figure}

\begin{figure}[htbp]	
	\centering
	\includegraphics[width = 0.9 \columnwidth] {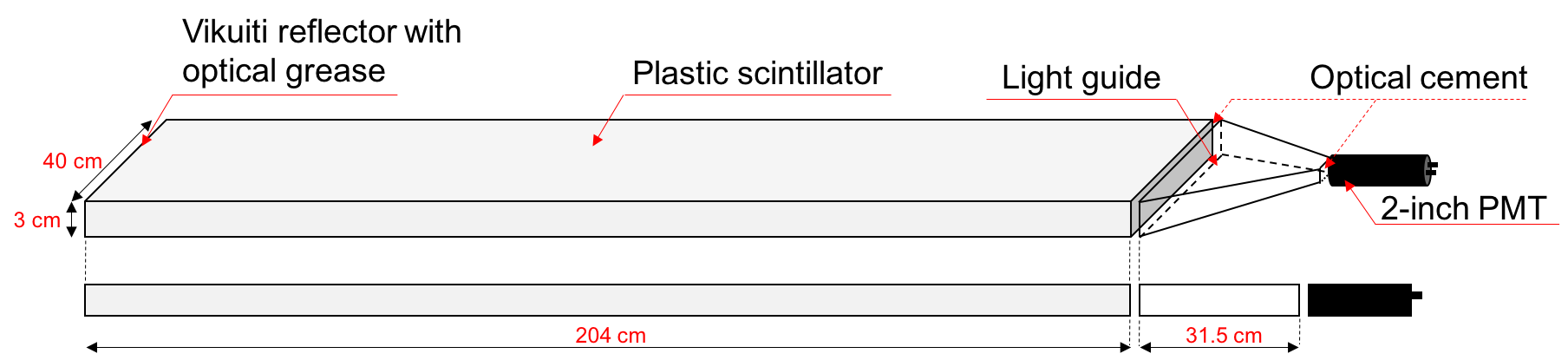}\\
	\caption{Schematic view of a right or left muon panel with assembly parts.}
	\label{detector}
\end{figure}

The muon detector for COSINE-100 consists of 37 panels of 3-cm thick Eljen EJ-200 plastic scintillator\footnote{http://www.eljentechnology.com}. Twenty seven t of the panels are 40~cm wide, ten are 33~cm wide in order to fit into the shielding structure. The array forms a near cubic structure with sides labeled as top, bottom, front, back, left, and right. The top-side panels are 282~cm long and are read out by a photomultiplier tube~(PMT) at both ends. The panels on the other sides are approximately 200~cm long and their signals are read out by one PMT. Table~\ref{tab:muonpannel} lists the dimensions of the muon panels for the six sides. Each panel is polished and coupled to an acrylic light guide using BC-600 optical cement from Saint-Gobain\footnote{http://www.crystals.saint-gobain.com}. Two-inch H7195 PMTs from Hamamatsu Photonics\footnote{http://www.hamamatsu.com} are mounted with optical cement at the end of the light guides. The optical coupling was visually inspected, as shown in Fig.~\ref{muonassemble}~(a). To increase the light collection, a Vikuiti reflector film is attached with optical grease at one end of the scintillator for the 200-cm-long panels. The muon panels are wrapped with a TYVEK reflector to collect light efficiently, as shown in Fig.~\ref{muonassemble}~(b). Then, the panel is covered with a 50-$\mu$m-thick aluminum foil~[Fig.~\ref{muonassemble}~(c)] and a black vinyl sheet~[Fig.~\ref{muonassemble}~(d)] to prevent external light from leakage inside and physical damage. A schematic of the muon panel used for the left or right side is shown in Fig.~\ref{detector}. 

\begin{table}[h!]
  \centering
  \caption{Plastic scintillator panels in the COSINE-100 experiment muon detector.}
  \label{tab:muonpannel}
  \begin{tabular}{c|ccccc}
				\hline
        Position & Length (cm) & Width (cm) & Thickness (cm) & Panels & PMTs\\
				\hline
        Top & 282 & 40 & 3 & 5 & 2\\
				\hline
        Front & 205 & 40 & 3 & 5 & 1\\
        & 207 & 33 & 3 & 2 & 1\\
				\hline
        Back & 202 & 40 & 3 & 5 & 1\\
        & 202 & 33 & 3 & 2 & 1\\
				\hline
        Right & 204 & 40 & 3 & 5 & 1\\
				\hline
        Left & 204 & 40 & 3 & 5 & 1\\
				\hline
        Bottom & 207 & 33 & 3 & 6 & 1\\
        & 205 & 40 & 3 & 2 & 1\\
				\hline
  \end{tabular}
\end{table}

\begin{figure}[htbp]
\centering
        	\includegraphics[width=.8\columnwidth]{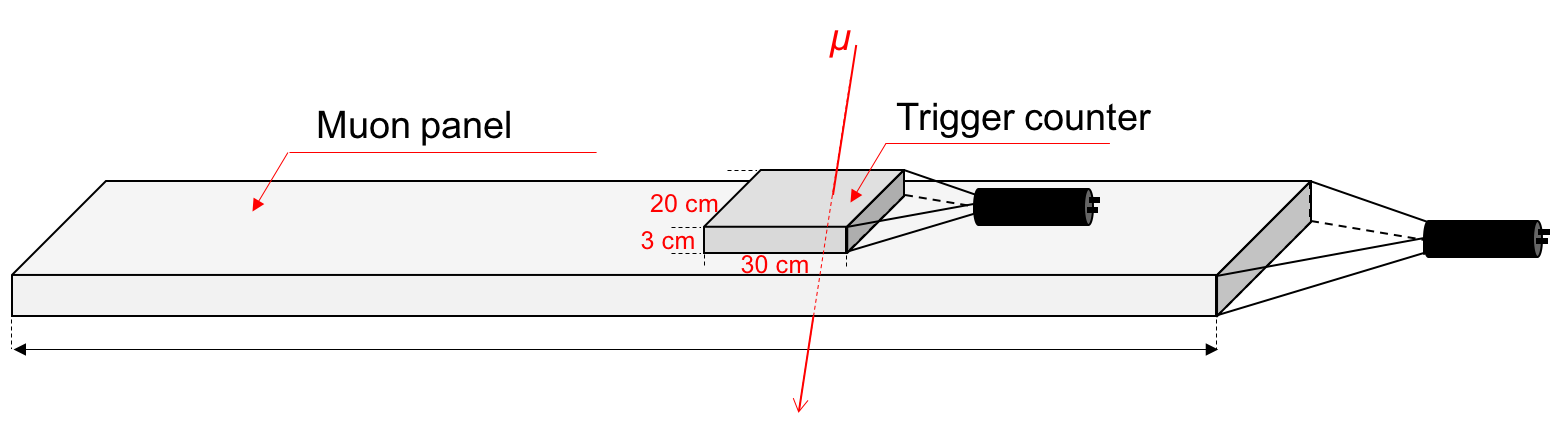}\\ 
        \caption{Experimental setup to test the performance of the muon panels.}
        \label{expsetup}
\end{figure}

\begin{figure}[htbp]
\centering
	\begin{tabular}{cc}
        	\includegraphics[width=.45\textwidth]{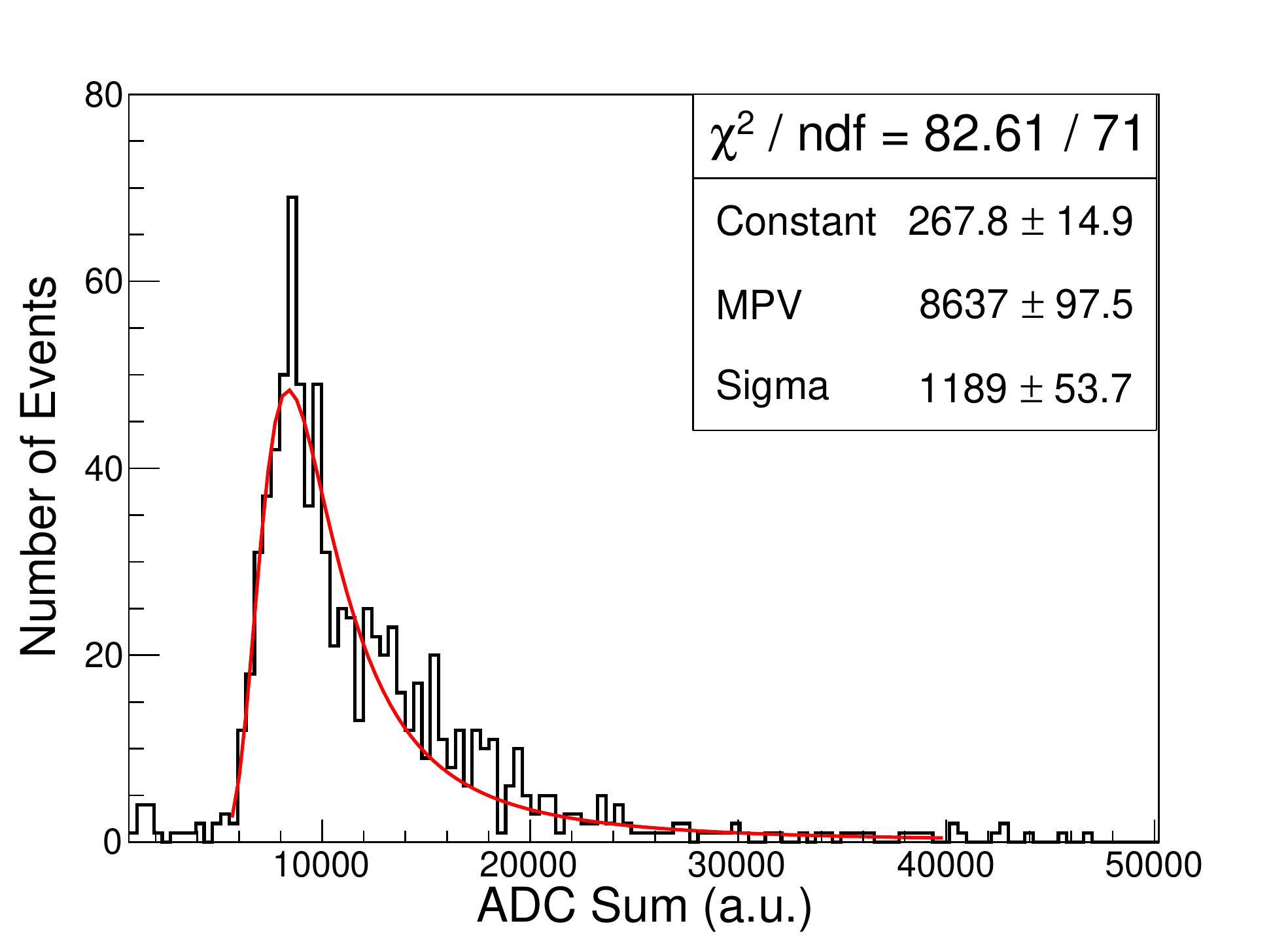} &
        	\includegraphics[width=.45\textwidth]{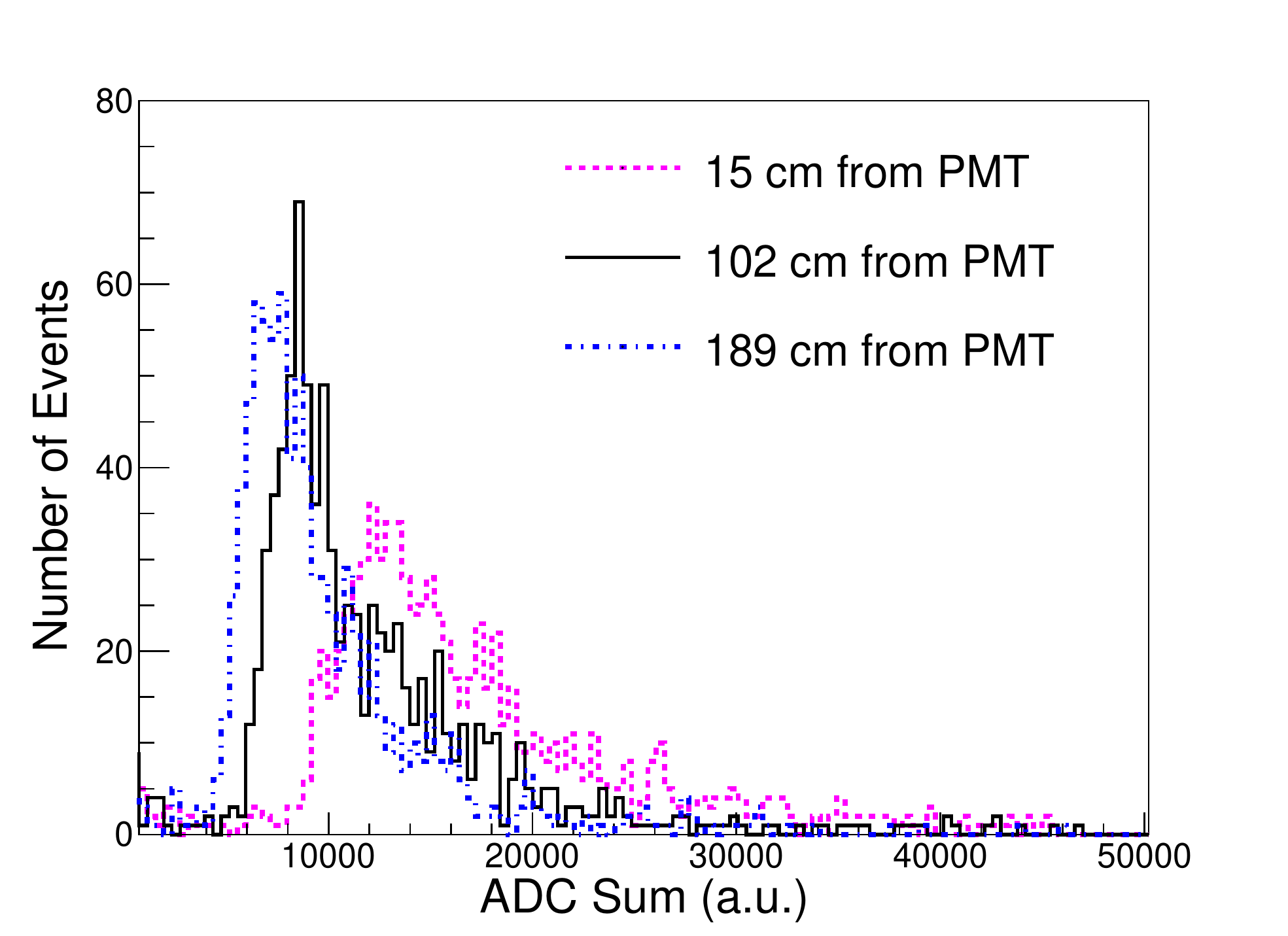} \\
					(a) & (b) \\
			\end{tabular}
        \caption{(a) Muon candidate events selected by the trigger counter and modeled by a Landau distribution. (b) Position-dependent charge distribution for a 204~cm long muon panel.}
        \label{landau}
\end{figure}

Various tests were performed on the muon panels in a ground laboratory where the cosmic ray muon flux is reasonably high. A small trigger counter made of the same plastic scintillator and placed above the muon detector panel as shown in Fig.~\ref{expsetup}. Coincident signals from the muon panel and the trigger counter are used to select muon candidates. With a sufficiently high energy threshold for the signals in the trigger counter, muon candidate events in the panel can be selected as shown in Fig.~\ref{landau}~(a). The muon candidate signals in the panels are well modeled by a Landau distributions.
The most probable value~(MPV) of a Landau fit to the panel signal distribution is used to estimate the relative light collection efficiency of each muon and found to be approximately 250 photoelectrons with 20\% panel to panel deviations. To study the position dependence of the light collection, the trigger counter was placed in three different positions: 15~cm~(the closest), 102~cm~(center), and 189~cm~(the farthest) from the PMT. As shown in Fig.~\ref{landau}~(b), the light collection strongly depends on the distance between the muon hit and the PMT; the maximum difference in the light collection was 45\%. However, muon candidate events far from the PMT were well separated from background events. The muon selection threshold for each panel is determined from the data accumulated at the lowest light yield locations. 

\begin{figure}[htbp]
\centering
        \includegraphics[width=.8\columnwidth]{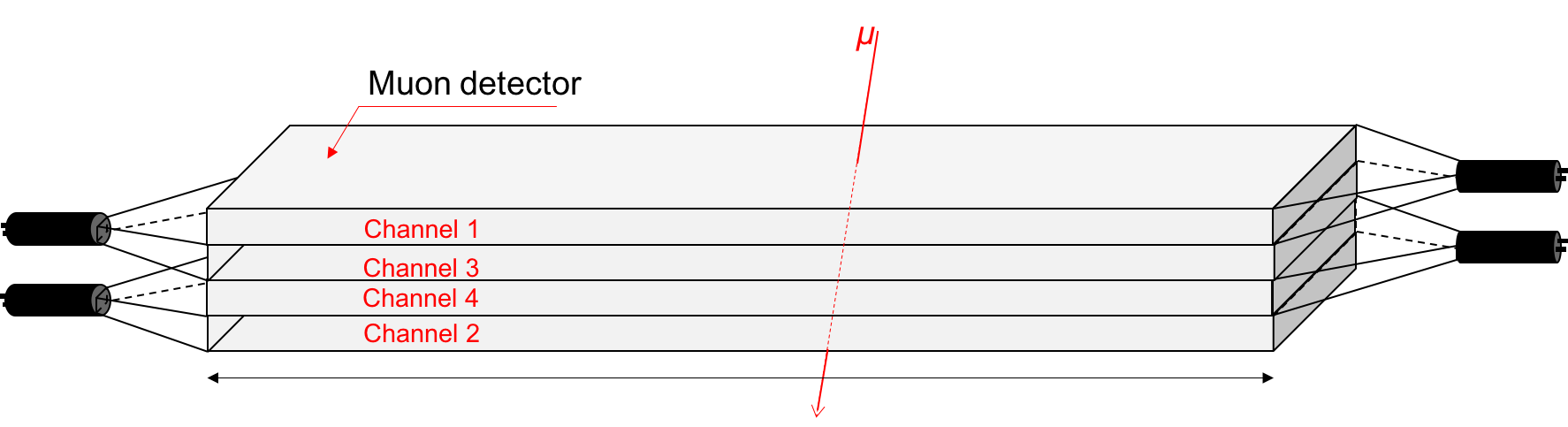}\\ 
        \caption{Schematic view of the muon tagging efficiency setup.}
        \label{muon_tag}
\end{figure}

\begin{figure}[htbp]
	\centering
	\begin{tabular}{ccc}
	\includegraphics[width = 0.33 \textwidth] {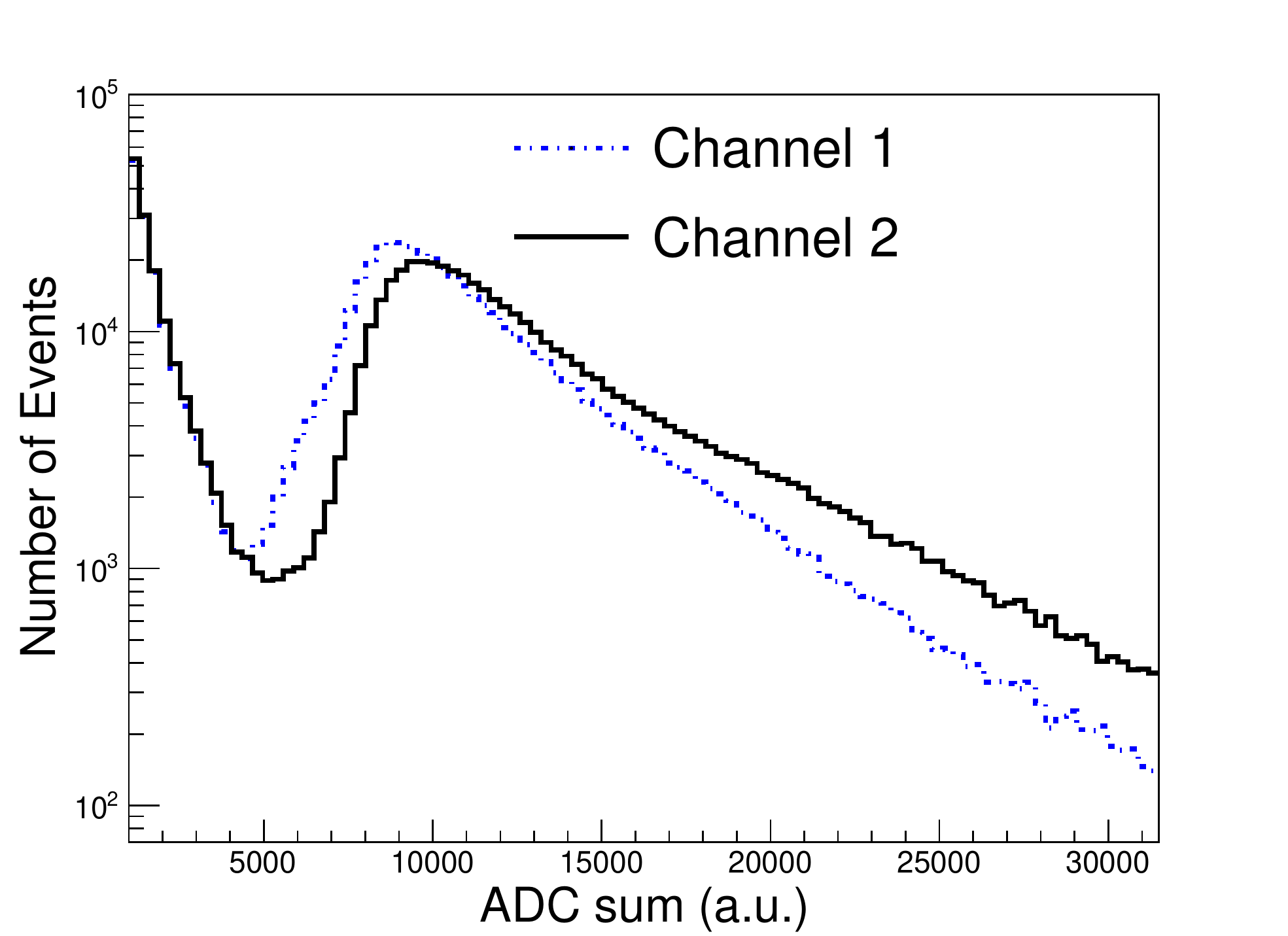}&
	\includegraphics[width = 0.33 \textwidth] {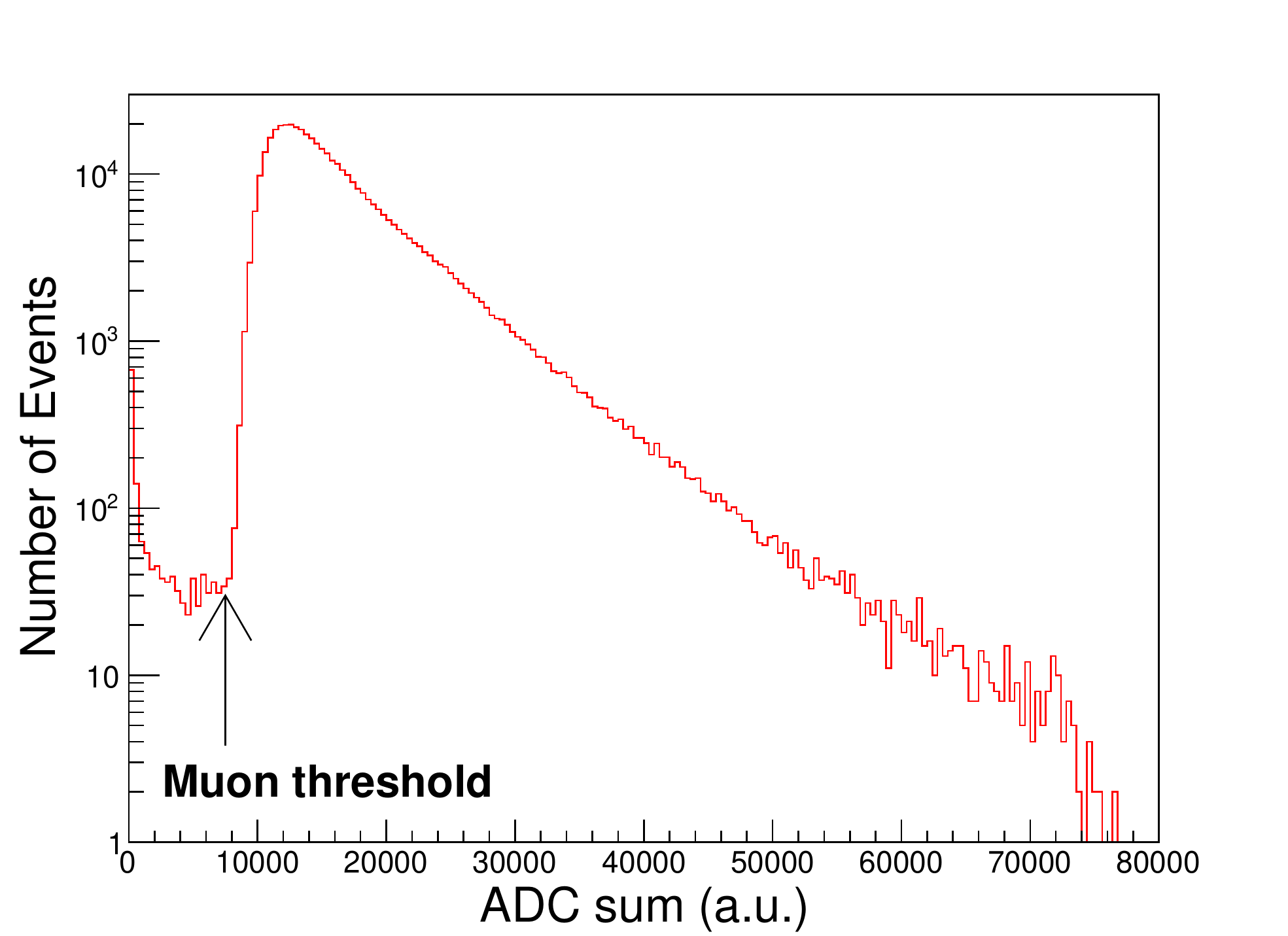}&
	\includegraphics[width = 0.33 \textwidth] {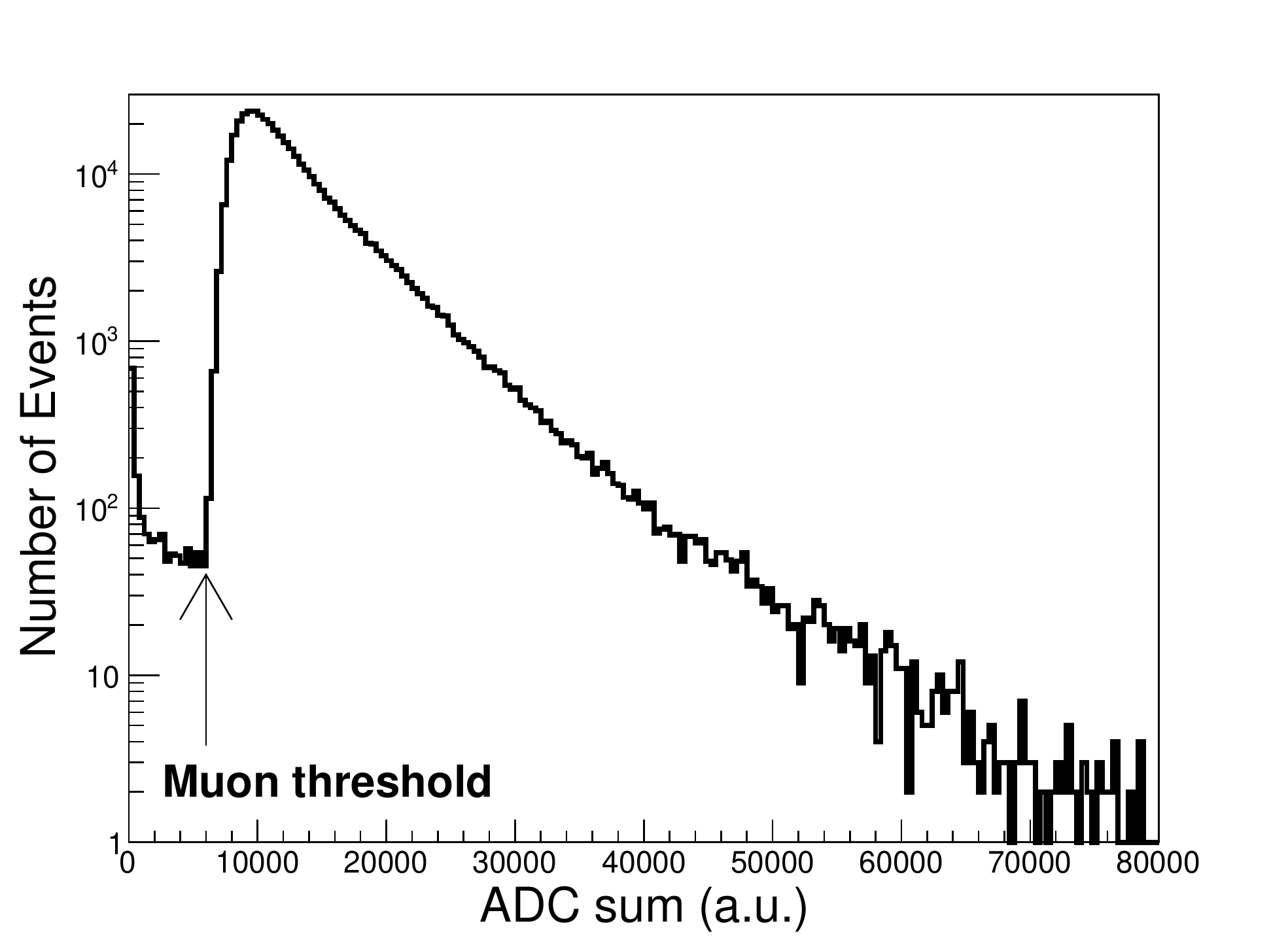}\\
	(a) & (b) & (c) \\
	\end{tabular}
	\caption{(a) Charge distributions of the two trigger panels in the setup of Fig.~\ref{muon_tag}. 
	Charge distributions of channel 3~(b) and channel 4~(c) applying muon selection requirements for channel 1~($>$4800) and channel 2~($>$5400).}
	\label{muoneff_ground}
\end{figure}

To estimate the muon detection efficiency as well as measure the muon flux, four panels were stacked as shown in Fig.~\ref{muon_tag}. 
A trigger for muon candidate events requires a coincident signals in a top~(channel 1)--bottom~(channel 2) pair. When the trigger is satisfied, data are recorded from all four channels. If muons passing through the top--bottom pair are selected, these muons should also pass the middle two panels. An offline selection signal threshold for muon candidates is applied for the top and bottom panels 
to remove the environmental $\gamma$--or $\beta$--induced backgrounds as shown in Fig.~\ref{muoneff_ground}~(a). 
For the charge distributions of the middle two panels, the selected events are inspected to see whether they are accepted as muon events 
as shown in Figs.~\ref{muoneff_ground}~(b) and \ref{muoneff_ground}~(c). A total of 425,824 events are selected as muon candidates from the top--bottom pairs. 
From this result, muon detection efficiency for the muon panels is determined approximately to be 99.6\%; the 0.4\% loss could include a systematic uncertainty due to minor misalignments of the panels. Muon candidate events selected by coincident signals from the middle two panels are used to determine the muon flux at the ground laboratory, which is 136~$\pm$~7~muons/m$^2$/s.

\section {Underground measurement}

\begin{figure}[htbp]
	\centering
	\begin{tabular}{cc}
	\includegraphics[width = 0.50 \textwidth] {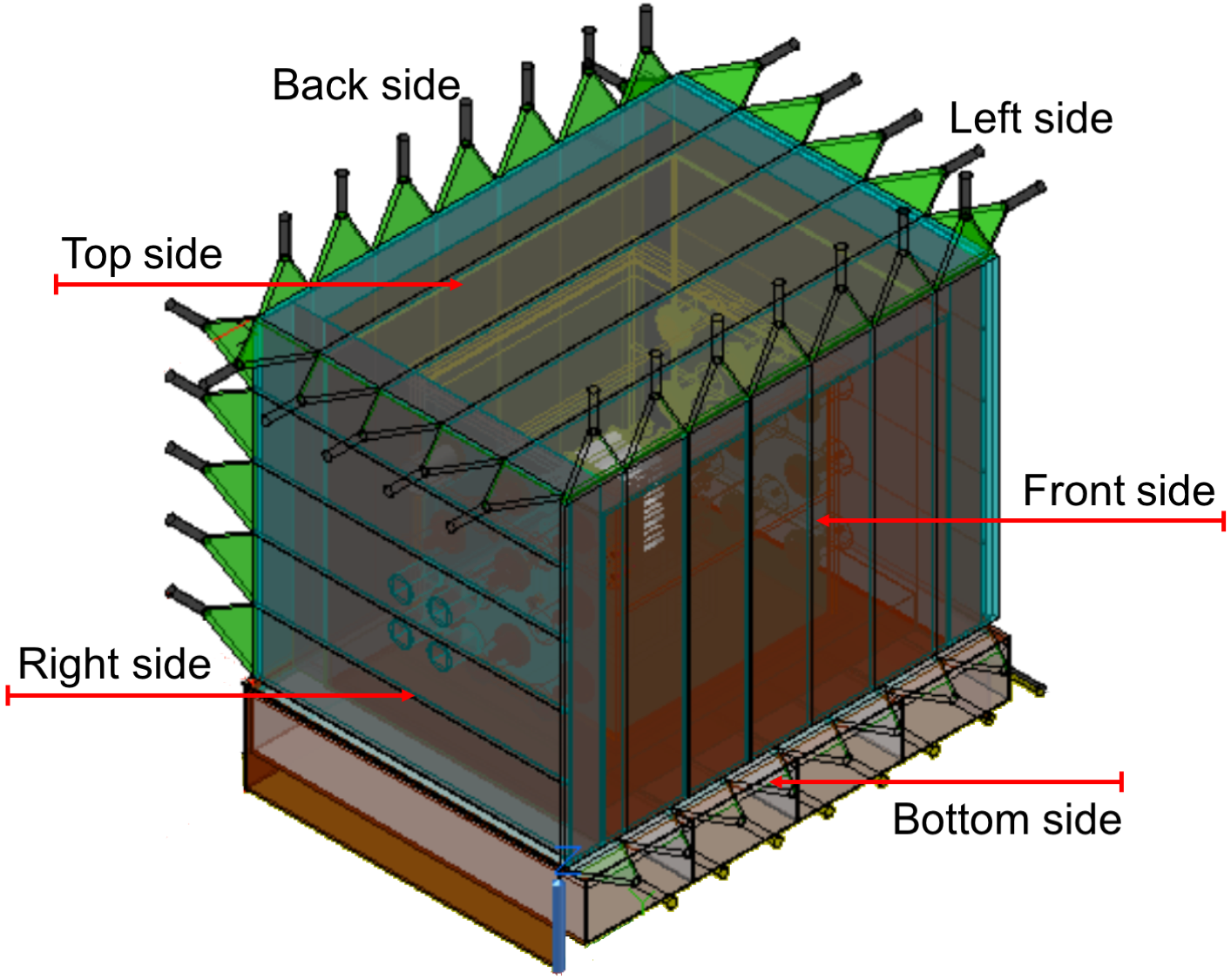}&
	\includegraphics[width = 0.46 \textwidth] {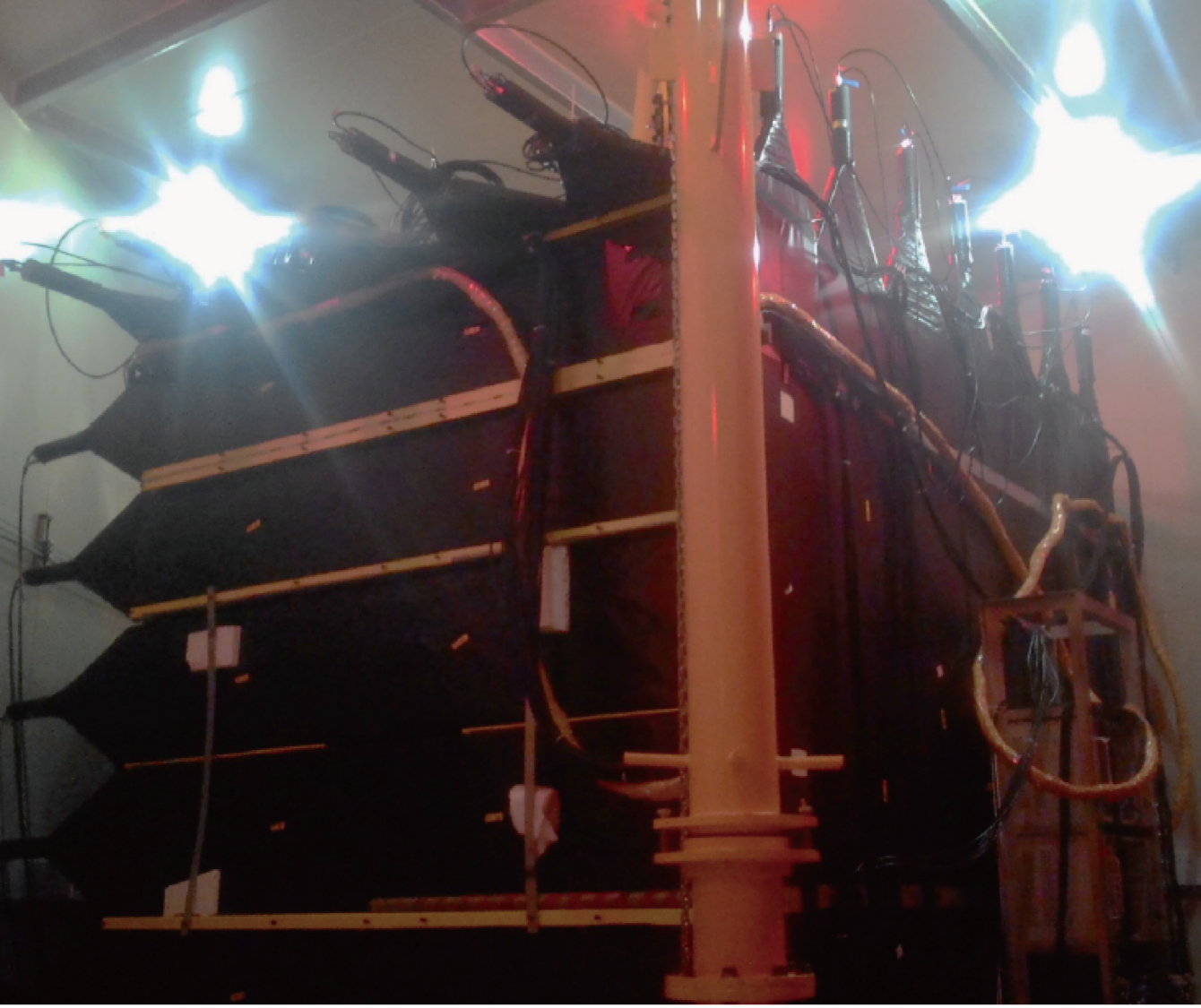}\\
	(a) & (b) \\
	\end{tabular}
	\caption{(a) Schematic of the COSINE-100 shielding structure, (b) Photograph of the installed muon detectors from front view.}
	\label{shielding}
\end{figure}

After the ground laboratory measurement, the muon panels were subsequently installed in COSINE-100 room at the Y2L as an active muon shield. The entire array consists of 37 panels that surround the detector with a full 4$\pi$ solid angle coverage. The general layout of the COSINE-100 muon detector and a photograph of the installed detector are shown in Fig.~\ref{shielding}. 

A total of 42 PMT signals are digitized by charge-sensitive 62.5 Megasamples per second ~(MS/s) ADCs from Notice Korea that are called~SADC\footnote{http://www.noticekorea.com}. One SADC module contains 32 channels with field-programmable gate arrays for signal processing and trigger generation. The dynamic range is 2 V with 12-bit resolution. Two modules are dedicated for muon detection. The SADC continuously calculates the integrated charges using a 192~ns integration time window with a 16~ns time bins. The trigger thresholds are set to be approximately one-third of the typical muon selection threshold. At least two PMTs from independent panels should exceed the trigger threshold at 4000~ADC within a 400~ns coincident time window. If this trigger condition is satisfied, each SADC channel records the maximum value of the 192~ns integrated charges and its time position within a 4~$\mu$s search window (a range of approximately $-$2 to 2$~\mu$s from the trigger position). Information for the six sides is calculated by combining the information of the panels constituting each side. The ADC charges are added for a given side panels, but the timing is selected from the one which has the maximum charge. 

Muon flux in underground laboratories shows significantly reduced rate~\cite{muonflux_underground}. Because the surviving deep underground muons still have high energies, most of the muons penetrate through all of the detector materials while leaving relatively large energy depositions. Energetic muons should deposit energies that are greater than the minimum ionization energy, which is approximately 6~MeV for a 3-cm-thick plastic scintillator~\cite{pdg} and is much larger than the typical $\gamma$ or $\beta$ energies produced by environmental background components. In addition, because the muon detector provides 4$\pi$ coverage of the COSINE-100 detector, muon signals can be expected to occur on at least two sides of the detector. Therefore, muon candidate events can be selected according to their energy deposits and coincidence requirements. 

\begin{figure}[htbp]
\centering
	\begin{tabular}{ccc}
        \includegraphics[width=.33\textwidth]{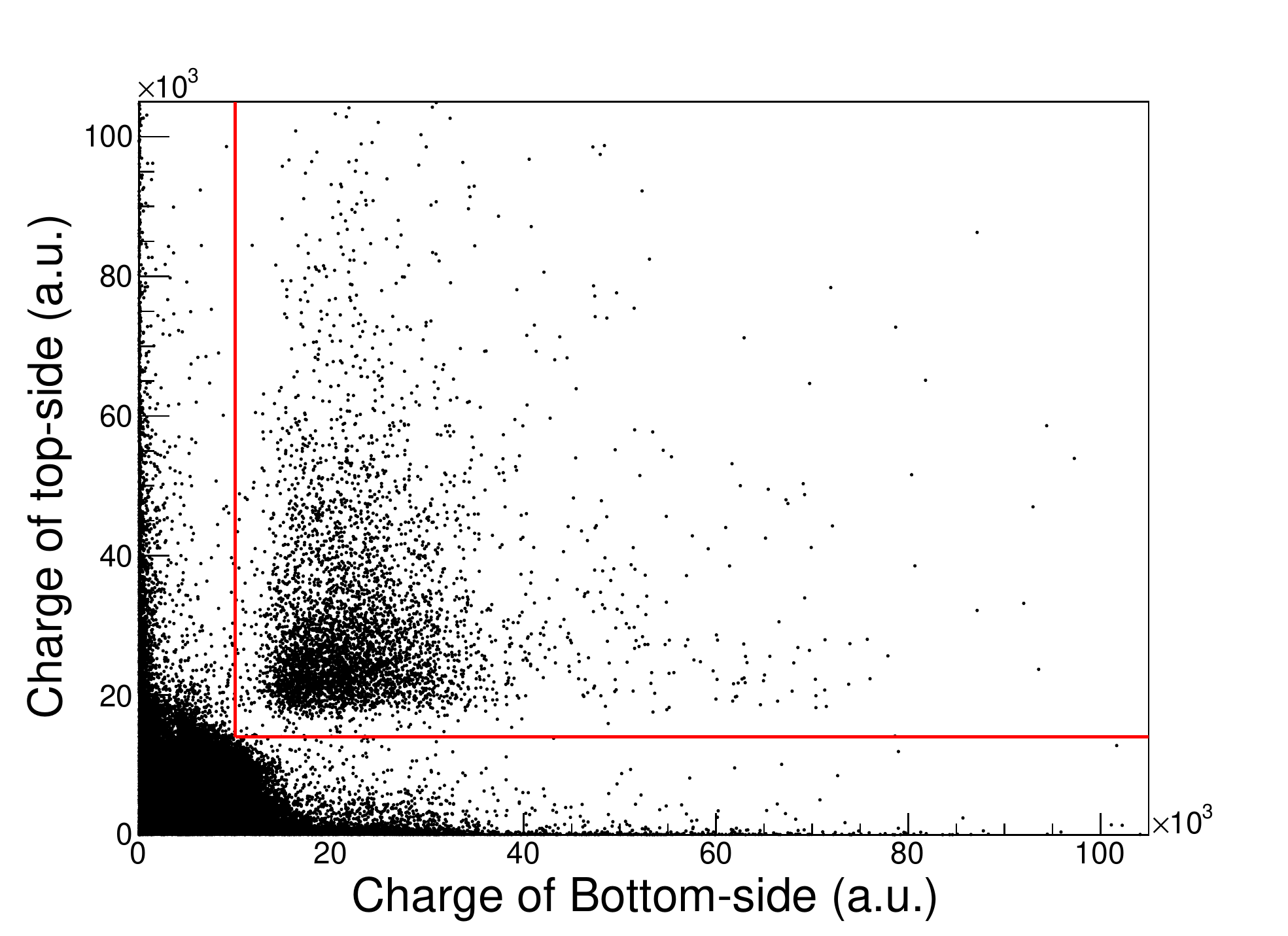} &
        \includegraphics[width=.33\textwidth]{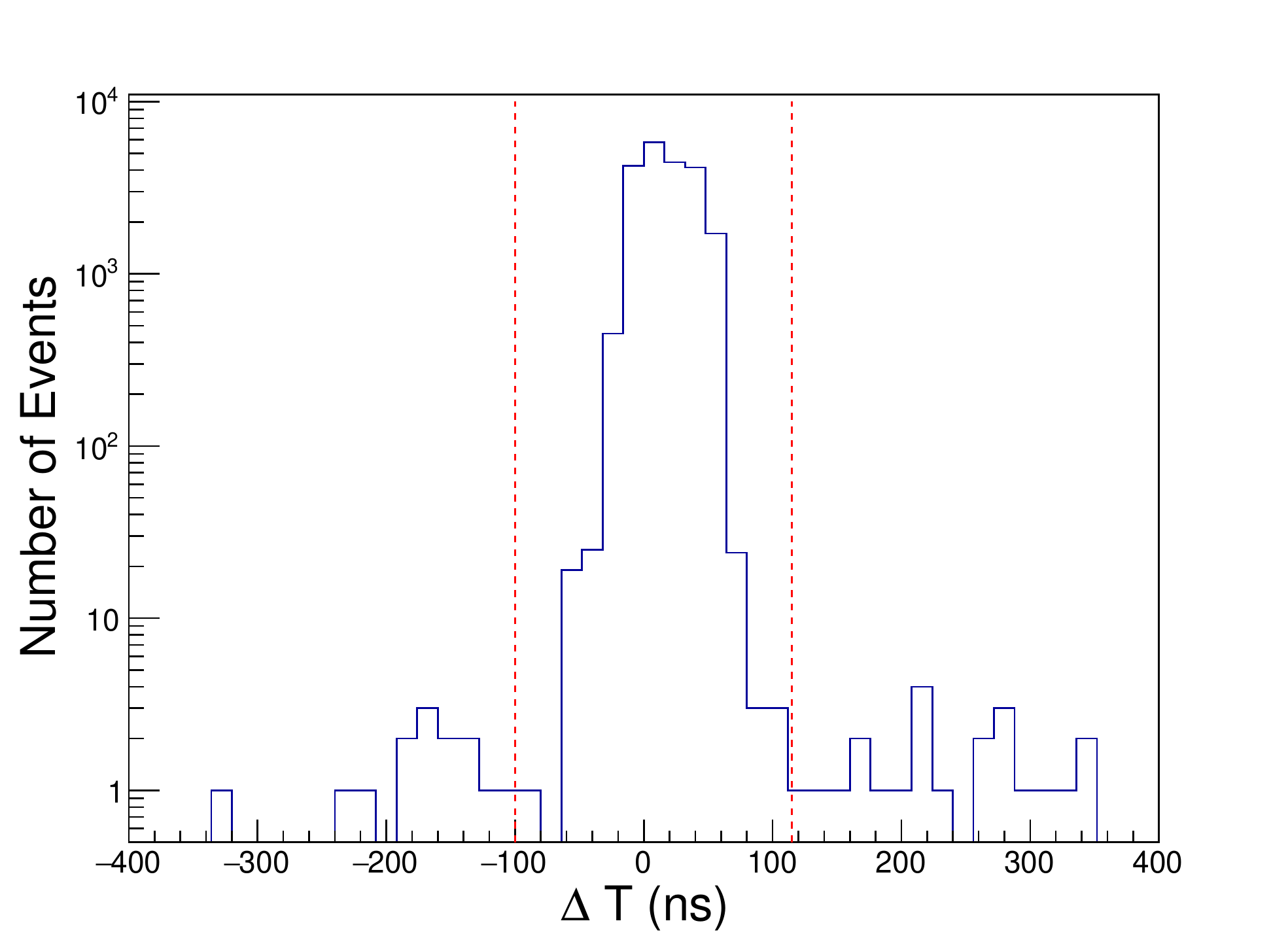} &
        \includegraphics[width=.33\textwidth]{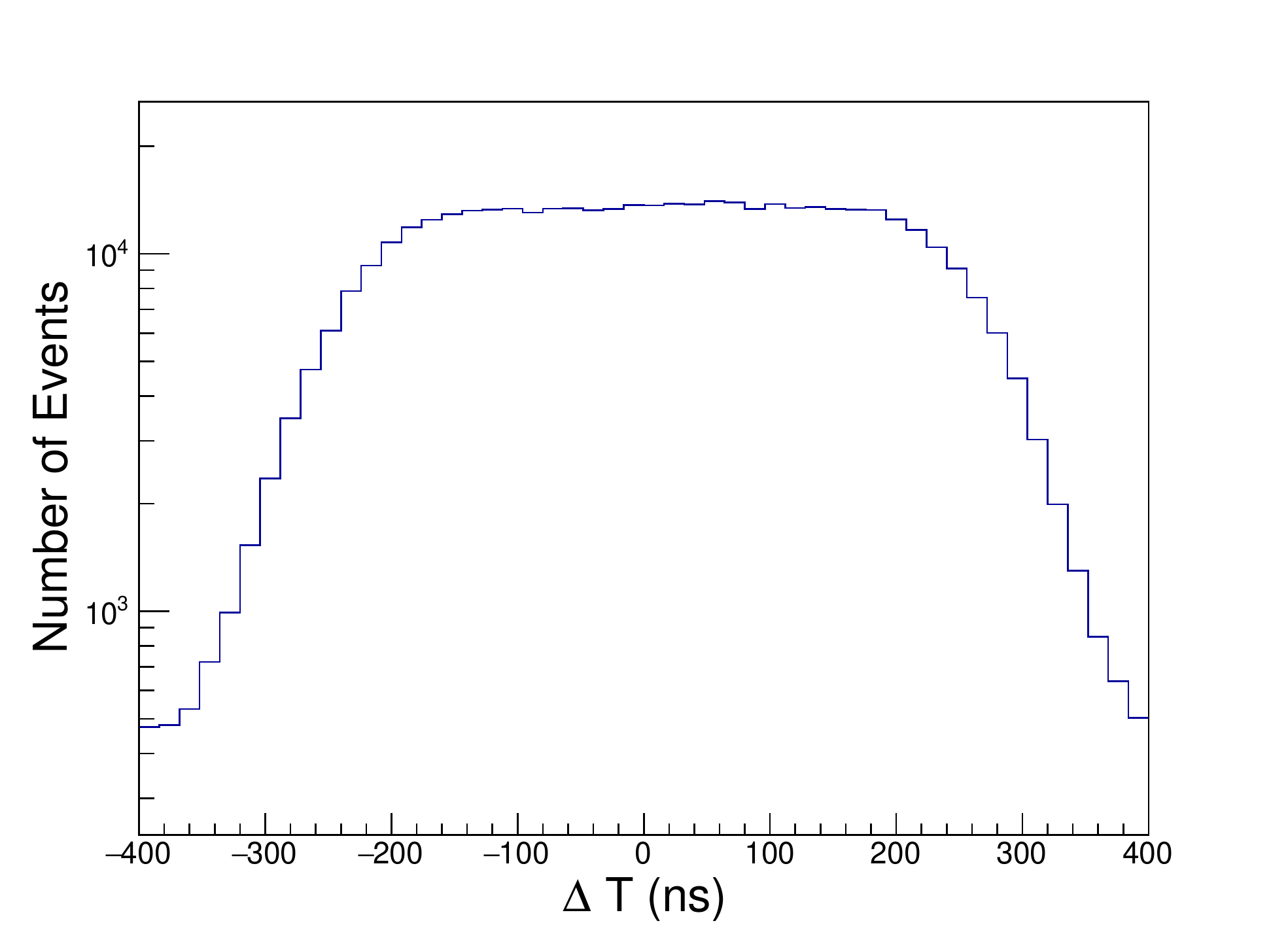} \\
				(a) & (b) & (c) \\
		\end{tabular}
        \caption{(a) A scatter-plot of the signal distributions for top- and bottom-side coincident events. The red lines represent the selection criteria for muon candidate events. Time differences~($\Delta$T) between top-side and bottom-side hits for muon candidate events~(b) and non-muon candidate events~(c).}
        \label{topcharge}
\end{figure}

The two-dimensional charge distribution of coincident events on two sides are analyzed. As an example, a scatter plot from the the top and bottom sides is shown in Fig.~\ref{topcharge}~(a). Time differences~($\Delta$T) between the bottom-side and top-side signals are shown in Fig.~\ref{topcharge}~(b), where muon candidate events exhibit a clear coincidence, while the $\gamma/\beta$ background events have a random distribution as shown in Fig.~\ref{topcharge}~(c). On the basis of the time correlation observed for the muon candidate events, an additional selection criterion of $\Delta$T is applied containing a 5$\sigma$ range of signal events, in this example $-100$~ns~<~$\Delta$T~<~115~ns as shown in Fig.~\ref{topcharge}~(b). 

Using the $\Delta$T distribution of the signal events, the background contamination in the signal region can be estimated. The number of events outside the signal region~($-200$~ns~$<\Delta$T~$<-100$~ns or $115$~ns~$<\Delta$T~$<215$~ns) is counted in Fig.~\ref{topcharge}~(b). Considering the background distribution in Fig.~\ref{topcharge}~(c), the background contamination in the signal region is calculated, which is 0.2\%. 

\begin{figure}[htbp]
\centering
        \includegraphics[width=.7\columnwidth]{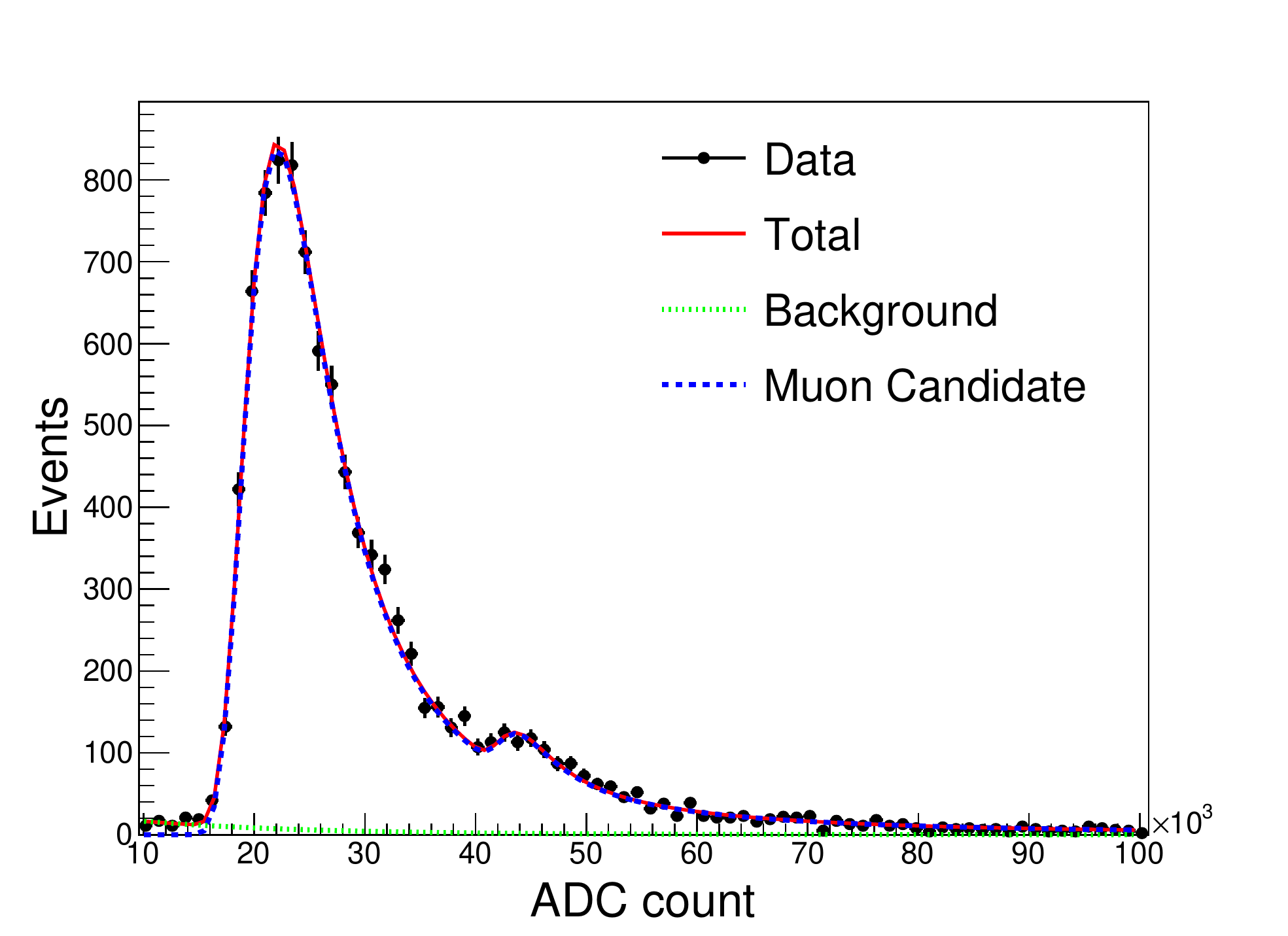}\\ 
        \caption{The charge spectrum of muon candidate events in the top-side panels with the charge threshold trigger requirement relaxed. An exponentially decaying background component and Landau-distributed muon signal component is used to model the data. Here we consider two Landau functions to take into account events hitting two panels, visible by the second peak around $44~\times~10^{3}$ ADC counts. The fit results indicate that the normally required threshold for the top-side charge rejects a negligible number of signal muon events~($<$0.1$\%$).}
        \label{landaufit}
\end{figure}

The muon selection criteria discussed above are applied to determine the charge distribution of the top-side panel. Here, the muon selection threshold for the top side is ignored to reveal the background shape, as shown in Fig.~\ref{landaufit}. Muon candidate events are fitted with the expected signal shape, Landau distributions, together with an exponential background component. Because the muons induce the spallation that makes multiple hits on the top-side, two Landau distributions are considered for signal modeling. From this fit, the background contribution in the signal region can be estimated as approximately $0.3\%$, which is consistent with the background contamination rate estimated with the $\Delta$T distribution. Furthermore, the muon selection efficiency was estimated to be 99.9$~\pm~0.1~$$\%$ when the charge threshold cuts mentioned previously are applied. A similar muon selection technique are applied for all pairs of different sides to tag muon candidate events.

The muon flux at the COSINE-100 experimental site is determined from all of two-side distribution studies that include signals in the top-side panel array. Muon events with hits in more than two sides are also counted as a single muon event. To normalize the muon rate, the effective area of the top-side is calculated, $A_d$ = 5.48~$\pm$~0.16~m$^{2}$, where the uncertainty reflects the area of the top-side panels that extend beyond the sides of the array and for which muon tagging is not fully active. This uncertainty is expected to be reduced with simulation-based studies.

\begin{figure}[htbp]
\centering
        \includegraphics[width=.9\columnwidth]{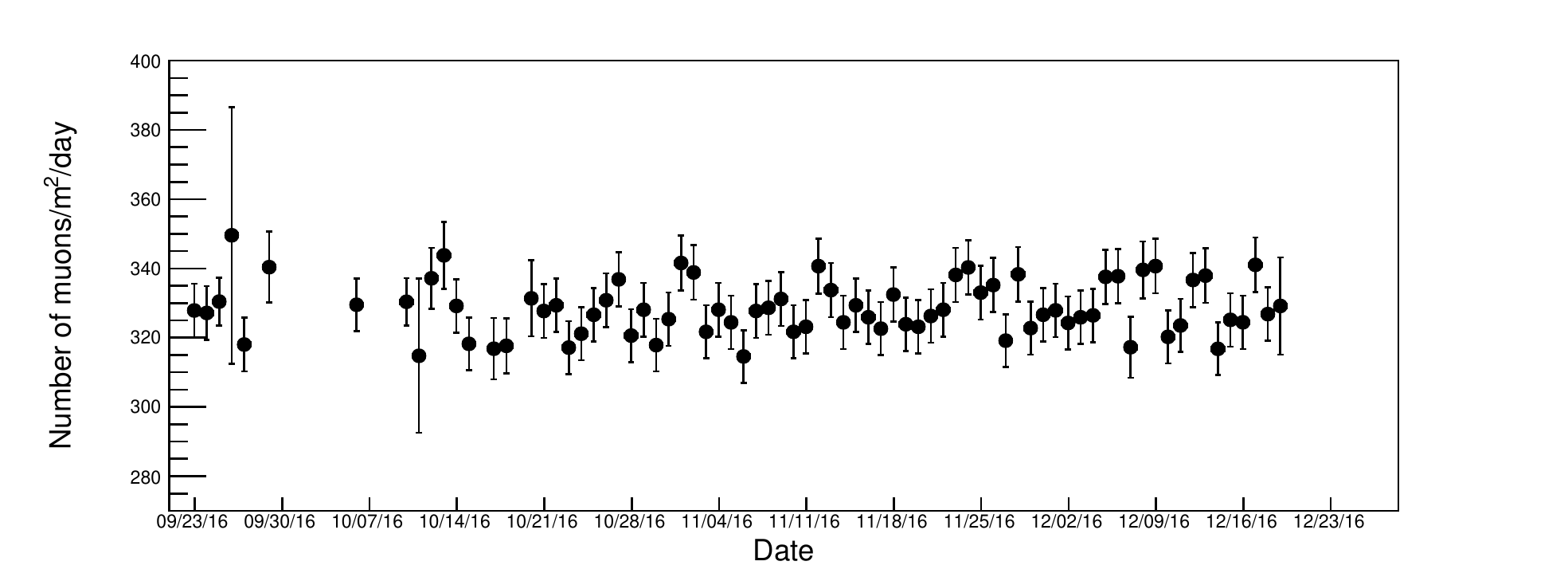}\\ 
        \caption{Measurement of the muon flux at the COSINE-100 detector over a three-month period.}
        \label{muonflux_y2l}
\end{figure}

A total number of $N_{\mu}$ = 144,235 muon candidate events were observed during a period of approximately three months of data acquisition. Because the non-muon contribution is negligible, the muon flux~$\Phi_\mu$~ can be calculated as follows:
\begin{equation}
\label{muonformula}
\Phi_\mu=\frac{N_{\mu}}{t_{d}\cdot A_{d}\cdot\epsilon_{\mu}},
\end{equation}
where $\epsilon_{\mu}$ is the muon selection efficiency, and $t_d$ is the total data acquisition time. This returns a measured muon flux at the COSINE-100 experimental site of 328$\pm$~1(stat.)$\pm$~10(syst.)~muons/m$^2$/day, where the systematic uncertainty is dominated by the top-side area calculation. This is slightly higher than that at the KIMS experimental site~\cite{kims_flux}, which is consistent with expectations based of surface geometry of Y2L and the approximately 200~m horizontal separation distance. This rate is approximately 2.8~$\times$10$^{-5}$ times the muon rate of the ground laboratory measurement. As shown in Fig.~\ref{muonflux_y2l}, the muon rate measured during a three-month period is very stable. A muon-induced event study with the NaI(Tl) crystals has started while data acquisition continues.

\section {Summary}
An array of muon detectors was installed around the outer detector shielding of the COSINE-100 experiment. The purpose is to tag and study the correlations between crystal and muon events that can mimic WIMP-like signals. Several tests were performed to validate the performance of the detector. 
The muon selection efficiency was found to be 99.9~$\pm~0.1$~\%, with only approximately 0.3\% fake muon events. 
This detection efficiency is sufficient for the COSINE-100 experiment. Using three months of data, the muon flux at the experimental site was measured to be 328$\pm$~1(stat.)$\pm$~10(syst.)~muons/m$^2$/day. Currently, data acquisition is ongoing to obtain sufficiently large statistics for the modulation analysis.

\acknowledgments
We thank the Korea Hydro and Nuclear Power (KHNP) Company for providing underground laboratory space at Yangyang. This work is supported by the Institute for Basic Science (IBS) under project code IBS-R016-A1, Republic of Korea; an Alfred P. Sloan Foundation Fellowship; NSF Grant Nos. PHY-1151795, PHY-1457995, DGE-1122492, and DGE-1256259; Wisconsin IceCube Particle Astrophysics Center (WIPAC), the Wisconsin Alumni Research Foundation; Yale University; and DOE/NNSA Grant No. DE-FC52-08NA28752, United States; STFC Grant ST/N000277/1, United Kingdom; and CNPq and grant \# 2017/02952-0, S\~{a}o Paulo Research Foundation (FAPESP), Brazil.

\end{document}